\documentclass[twocolumn]{aastex61}
\usepackage{color}
\usepackage{verbatim}
\usepackage{amsmath}

\shorttitle{Single Transits in TESS}
\shortauthors{Villanueva Jr., Dragomir, \& Gaudi}

\begin{document}

\title{An Estimate of the Yield of Single-Transit Planetary Events from the Transiting Exoplanet Survey Satellite}

\email{svillan@astronomy.ohio-state.edu}
\author{Steven Villanueva Jr.}
\affil{Department of Astronomy, The Ohio State University, 140 West 18th Av., Columbus, OH 43210, USA}

\author{Diana Dragomir}
\affil{Massachusetts Institute of Technology, Cambridge, MA 02139 USA}
\affil{Hubble Fellow}

\author{B. Scott Gaudi}
\affil{Department of Astronomy, The Ohio State University, 140 West 18th Av., Columbus, OH 43210, USA}

\submitjournal{AAS Journals}

\newcommand{\myemail}{svillan@astronomy.ohio-state.edu}

\begin{abstract}
We present a semi-analytic estimate of the expected yield of single-transit planets from the Transiting Exoplanet Survey Satellite (TESS).  We use the TESS Candidate Target List 6 (CTL-6) as an input catalog of over 4 million sources. We predict that from the 200,000 stars selected to be observed with the high-cadence postage stamps with the highest CTL-6 priority, there will be 241 single-transit events caused by planets detectable at a signal-to-noise ratio of SNR$\ge7.3$. We find a lower limit of an additional 977 events caused by single-transit planets in the full frame images (FFI); this is a lower limit because the CTL-6 is incomplete below a TESS magnitude of $T>12$. Of the single-transit events from the postage stamps and FFIs, 1091/1218 will have transit depths deeper than 0.1\%, and will thus be amenable for photometric follow-up from the ground, and 1195/1218 will have radial velocity signals greater than 1~m/s. We estimate that the periods of 146 single transits will be constrained to better than 10\% using the TESS photometry assuming circular orbits. We find that the number of planets detected by TESS in the postage stamps with periods $P>25$~days will be doubled by including single-transiting planets, while the number of planets with $P>250$~days will be increased by an order of magnitude. We predict 79 habitable zone planets from single-transits, with 18 orbiting FGK stars.
\end{abstract}


\section{Introduction}

The Transiting Exoplanet Survey Satellite (TESS), launched in Spring 2018, will discover thousands of transiting exoplanets that exhibit two or more transits during the mission. TESS will have a number of advantages over previous transiting planet surveys, including ground-based surveys such as the Hungarian Automated Telescope \citep{bakos04} survey, the Wide Angle Search for Planets \citep{pollacco06} survey, and the Kilodegree Extremely Little Telescope \citep{siverd12} survey, as well as space-based missions like Corot \citep{baglin03}, Kepler \citep{borucki10}, and K2 \citep{howell14}. Ground-based surveys are essentially limited to planets whose transits have depths above $\sim0.1\%$, but do so around bright stars that are amenable to follow-up observations from ground-based telescopes and radial velocity measurements. Kepler has exquisite photometric precision down to several tens of parts-per-million (ppm) \citep{koch10}, but the majority of the planets found by Kepler are orbiting stars that are too faint to be confirmed via radial velocity using the current generation of telescopes and instruments. Both the original Kepler campaign and the extended K2 mission campaigns have relatively long baselines of almost 4 years and 80 days respectively, but both are also limited in their sky coverage.

By virtue of TESS's observing strategy and design, it will observe $85\%$ of the entire sky, monitoring and discovering planets transiting bright stars, which are amenable to both photometric and radial velocity follow-up, as well as detailed characterization of their atmospheres via ground and space-based telescopes. The trade-off of achieving this nearly all-sky coverage is that $63\%$ of the sky, or $74\%$ of the mission's total sky coverage, will only be observed for 27 days (as compared to 80 days for each K2 campaign and nearly 4 years for the primary Kepler campaign).  In this regime of many millions of stars monitored for a relatively short amount of time, the number of single-transit planetary events found by TESS will be much larger than that expected or found by Kepler \citep{yee08,dfm16}.  Single transit events require significantly more resources to confirm than planets that exhibit two or more transits, but nevertheless can be quite scientifically valuable. 

With the planned survey strategy, and a requirement of at least two transits to confirm a planet, the majority of these planets will have periods of less than 10 days. The primary mission of the survey is to measure masses and radii of 50 terrestrial planets. This leaves open the opportunity to discover planets outside of this regime, including giant planets and planets on long orbits that transit only once. However, recovering, confirming and studying planets that transit only once pose difficulties.  Some of these difficulties include the fact that their ephemerides are difficult to constrain for the purpose of scheduling of future observations, and they are easily confused with false-positives.

Previous studies have investigated single-transit events in Kepler \citep{yee08}. Multiple simulations have been performed to estimate the (two or more transit) yield of TESS \citep{sullivan15,bouma17,ballard18,barclay18}. In each of these simulations, only systems that exhibit two or more transits are reported, and to date there has not been an estimate of the expected yield of single-transits in TESS, or their properties. Given the number of stars and observing strategy of TESS, we expect a 100 fold increase in the number of single-transit events in TESS relative to Kepler.

It is both worthwhile and possible to follow up these longer-period transiting planets, as they represent an opportunity to investigate a number of questions related to planet formation, such as the migration mechanism for Hot Jupiters, and the physical mechanisms that lead to inflated radii of close-in giant planets.  Using the definition of habitable zones described by \cite{kopparapu13}, transiting planets of main sequence stars of spectral type earlier than roughly M5 ($T_{\mathrm{eff}}\approx2800$~K) will have the inner edge of the habitable zone at periods of $\approx11$ days. For the majority of the TESS survey, any habitable zone planets around M4 or earlier stars are expected to only display single transits.

\section{Expected Number of Single-Transit Planets}\label{sec:singles}

The expected total number of planets detectable by TESS with exactly one or more transits is the integral over all periods and all planetary radii of the geometric probability of detecting a transit around a star $\wp_{tr}$, the probability of observing the transit(s) during the finite baseline of observations $\wp_{B}$, and the planet occurrence rates $f(P)$ with a Heaviside step function cut on the signal-to-noise ratio $\Theta(\Delta SNR)$, multiplied by the total number of stars observed by TESS $N_{\star}$
\begin{eqnarray}\label{ep:yield}
    N_{det}= N_{\star}\int\wp_{tr}\wp_{B}f(P)\Theta(\Delta SNR)\mathrm{d}P
\end{eqnarray}
where $\Delta SNR = SNR-SNR_{\mathrm{min}}$, where each term is for a fixed period and planetary radius. In reality, each of the terms [$N_{\star}$, $\wp_{tr}$, $\wp_{B}$, $f(P)$, $\Theta(\Delta SNR)$] depend on more than just the period $P$ and planet radius, but also depend on other variables such as the stellar mass, stellar radius, apparent stellar magnitude, and intrinsic stellar variability. All of these variables are considered in the final analysis.

We will also assume circular orbits and that $r_p\ll R_{\star}$.  With this assumption, the geometric transit probability is then
\begin{eqnarray}
    \wp_{tr}=\frac{R_{\star}}{a}
\end{eqnarray}
for non-grazing geometries, where $R_\star$ is the host star radius. It is possible to evaluate Equation~\ref{ep:yield} in units of the semi-major axis $a$, but it is more convenient to use the period $P$ as this is the direct observable in both transit and radial velocity detections of exoplanets. We use Kepler's third law assuming that the planet's mass is much smaller than the stellar mass to convert semi-major axis to period
\begin{equation}\label{eq:kepler}
    P^2=a^3\frac{4\pi^2}{G(M_\star+m_{p})}\approx a^3\frac{4\pi^2}{GM_\star}
\end{equation}
where $M_\star$ is the stellar mass. The geometric transit probability then becomes a function of stellar mass, stellar radius, and orbital period:
\begin{eqnarray}\label{eq:geo}
    \wp_{tr}=\left(\frac{4\pi^2}{G}\right)^{1/3}R_{\star}M_\star^{-1/3}P^{-2/3}
\end{eqnarray}
The geometric probability decreases as $\wp_{tr}\propto P^{-2/3}$ and leads to a decreased probability of detection at long periods.

We also consider the probability of a transit occurring during the finite baseline of observation $B$ of the TESS mission $\wp_B$. During this paper, we will evaluate cases where two or more transits occur, or exactly one transit occurs. In the case where both two or more transits are observed with planets on periods shorter than $B/2$, the probability of observing two or more transits occurring during the observing baseline is unity. However for planets on periods longer than $B/2$, the probability decreases until only one, or no transits occur during the observing baseline. For a finite observing baseline $B$ and ignoring the finite duration of the transits, the probability of exactly one $\wp_{B,1}$, or two or more $\wp_{B,2+}$ transits occurring is given by:
\begin{eqnarray}\label{eq:baseline}
    \begin{array}{rll}
        \wp_{B,1}&=\frac{B}{P}&,P\ge B\\
        &=\frac{2P}{B}-1&,\frac{B}{2}\le P\le B\\
        \wp_{B,2+}&=2-\frac{2P}{B}&,\frac{B}{2}\le P\le B\\
        &=1&,P\le \frac{B}{2}
    \end{array}
\end{eqnarray}
These are analogous to Equations 3 and 4 from \cite{yee08}, but we include the case of two or more transits, instead of exactly two transits. \cite{yee08} incorrectly chose the lower limit for the two transit case to be $P/4$, instead of $P/3$, however, this only biases their yields for the two-transit cases and their single-transit yields should be unaffected.
When investigating single-transit events with periods longer than the observing baseline, the total probability is
\begin{eqnarray}
    \wp_{tot}&\equiv&\wp_{tr}\wp_B\nonumber\\
    &=&0.026\left(\frac{R_\star}{R_\odot}\right)\left(\frac{M_\star}{M_\odot}\right)^{-1/3}\left(\frac{P}{27.4\mathrm{d}}\right)^{-5/3}\left(\frac{B}{27.4\mathrm{d}}\right)
\end{eqnarray}
and scales as $\wp_{tr}\wp_B\propto P^{-5/3}$.

\begin{figure*}[t]
\centering
\includegraphics[width=8cm]{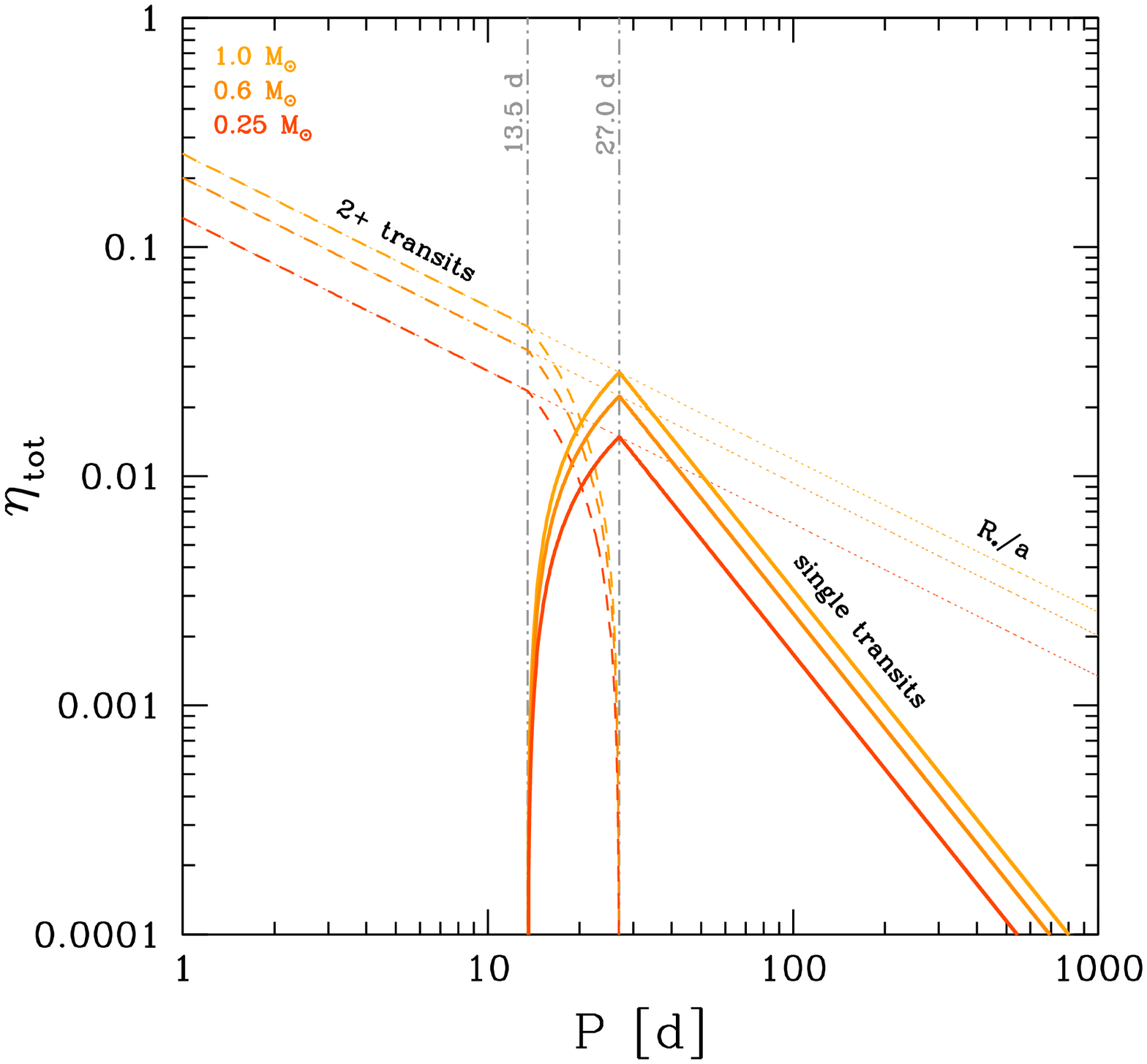}
\includegraphics[width=8cm]{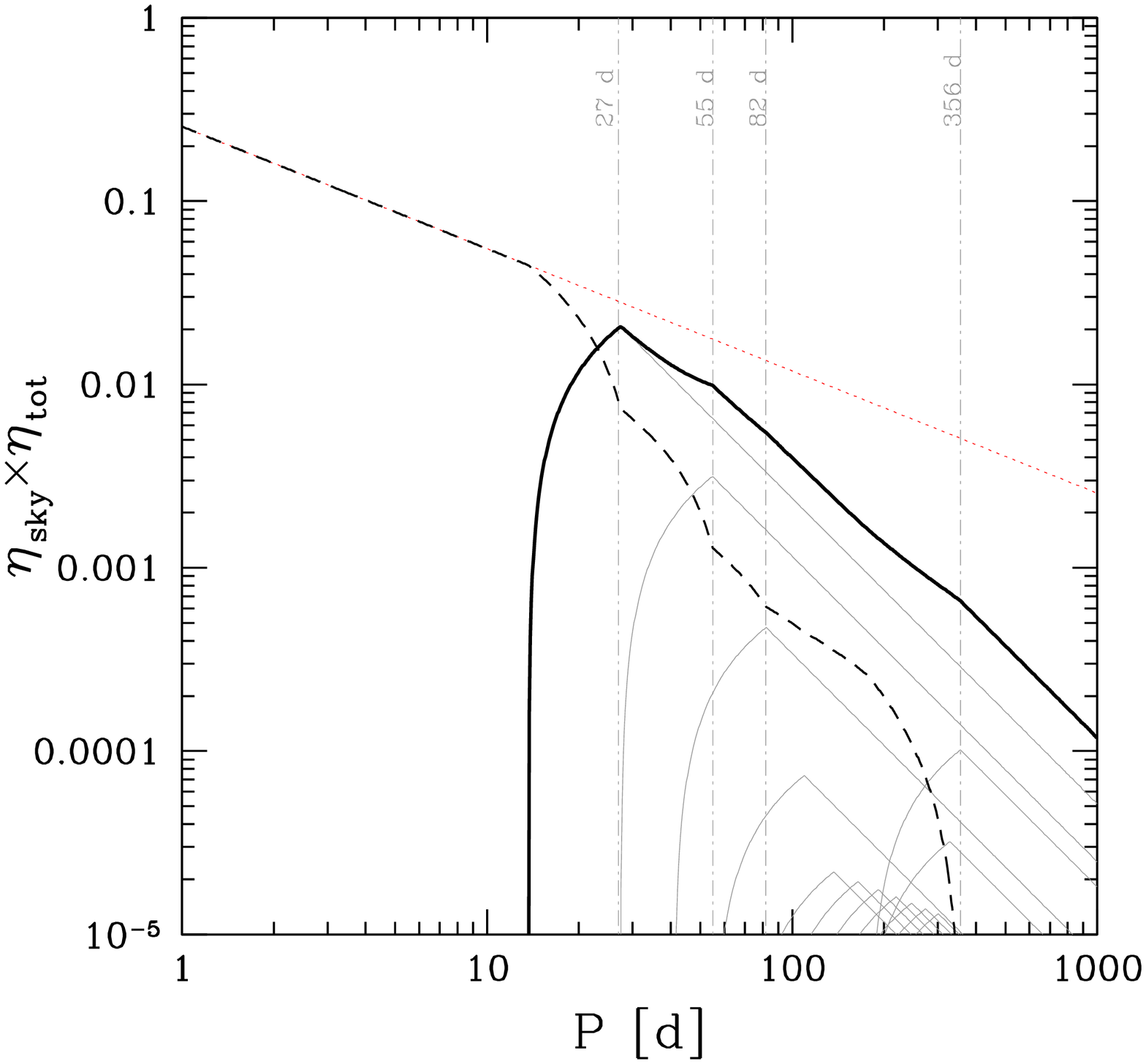}
\includegraphics[width=8cm]{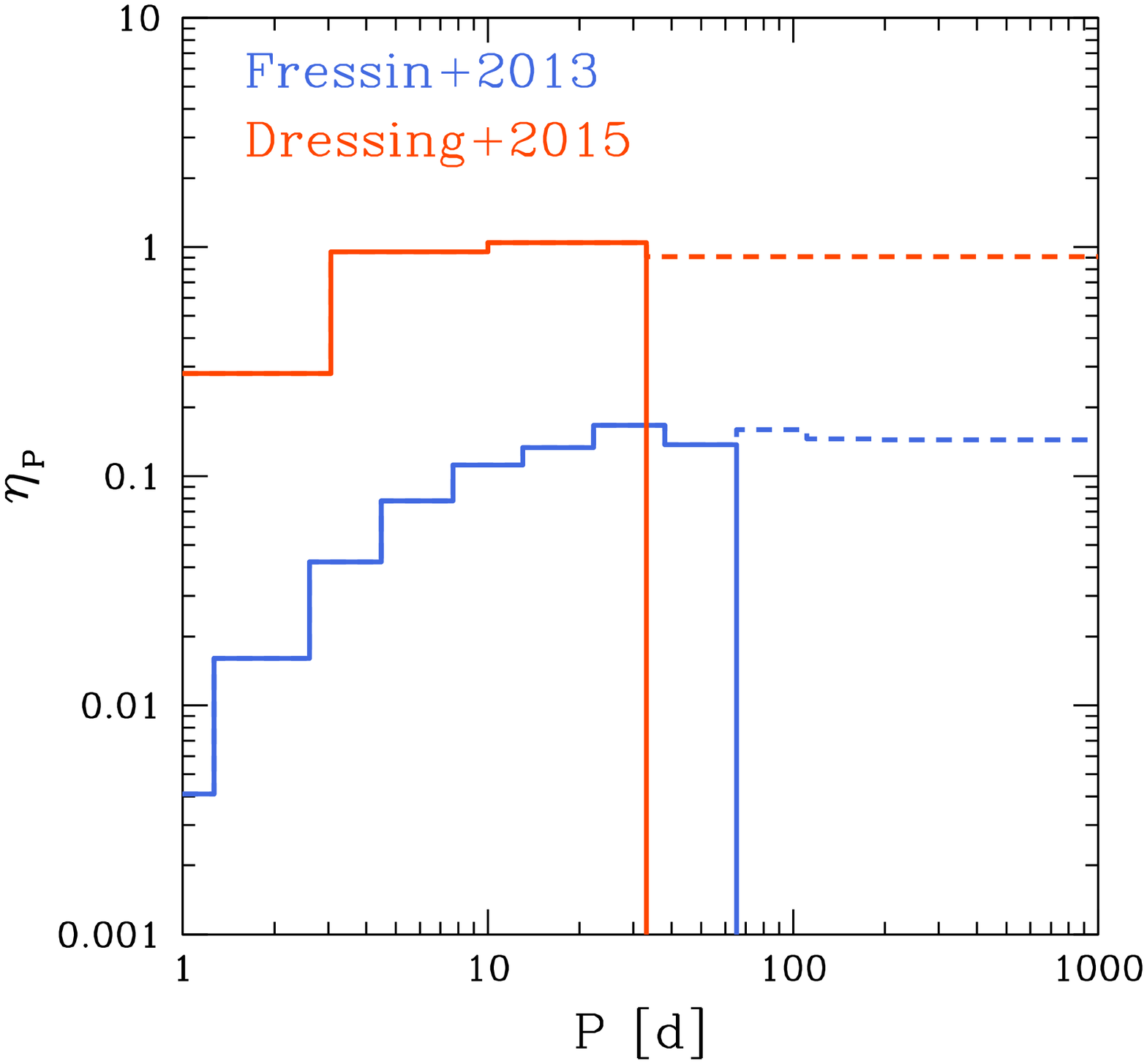}
\includegraphics[width=8cm]{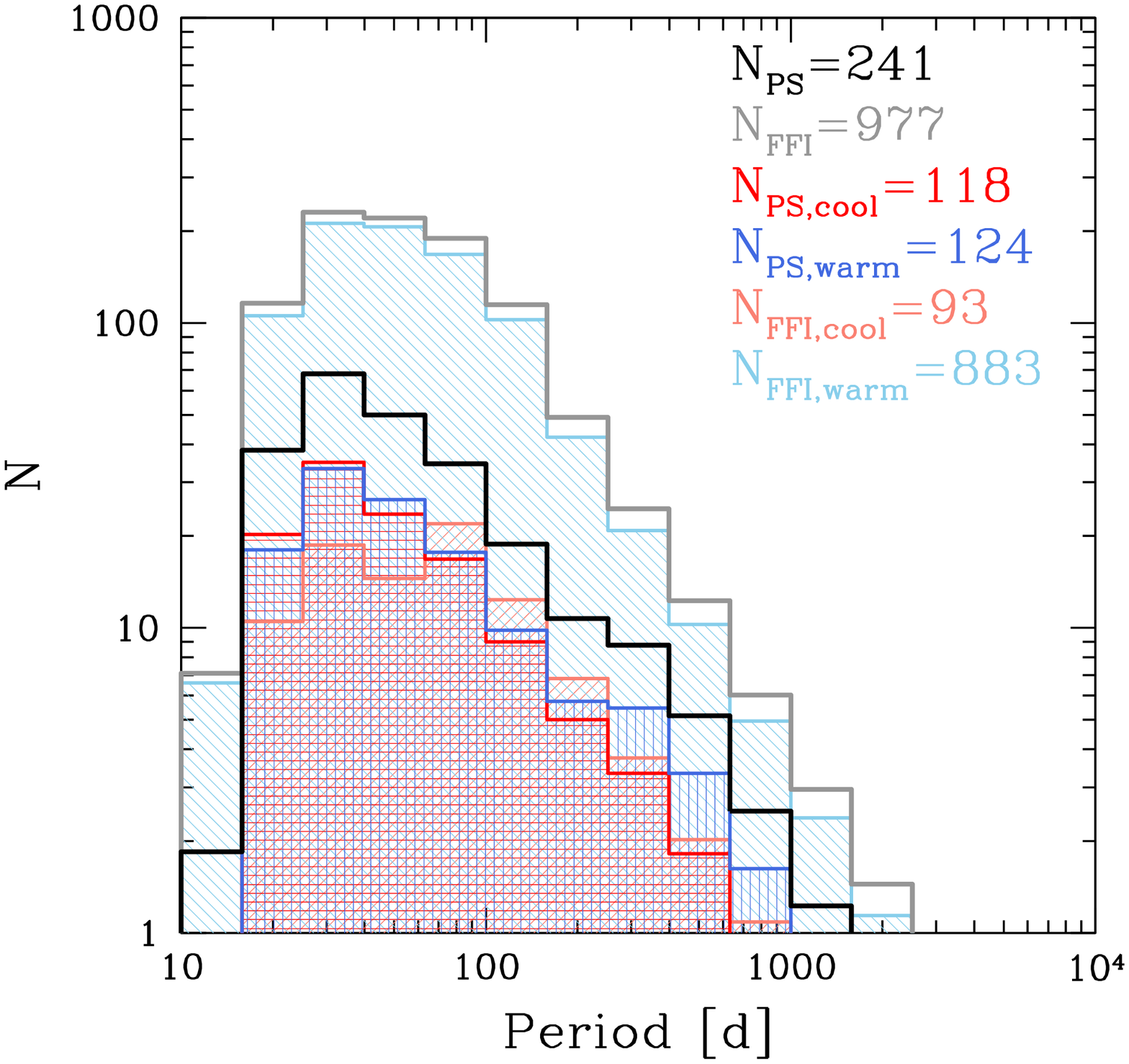}
\caption{\label{fig:yield}\textit{Top Left:} Probability of observing a single transit (solid) or two or more transits (dashed) for the 27.4 day baseline as compared to the geometric transit probability (dotted). Colors correspond to a 1.0, 0.6, and 0.25 $M_{\odot}$ host star. \textit{Top Right:} Mission-weighted probability of observing a single transit (solid black line) or two or more transits (dashed black line) over all observing baselines as compared to the geometric transit probability (red dotted line) for a 1.0 $M_{\odot}$ host star. All 13 individual single transit probability curves, corresponding to the 13 possible baselines, are shown as grey solid lines for reference. \textit{Bottom Left:} Planet occurrence rates for $T_{\mathrm{eff}}\ge4000$~K stars (blue) and $T_{\mathrm{eff}}<4000$~K stars (red). Dark lines are the fraction of stars expected to host at least one planet in each period bin, while the dotted lines represent our extrapolation to long periods. \textit{Bottom Right:} Combining the first three panels with the total number of stars, we estimate the period distribution of single transit events expected from TESS postage stamps (black) and in the FFIs (gray). Events around stars with $T_{\mathrm{eff}}\ge4000$K are in blue, and stars $T_{\mathrm{eff}}<4000$~K in red. The darker shades are for the 2-minute cadence, while the lighter shades are for the 30-minute cadence. The total number of planets exhibiting a single transit event expected from the TESS mission is over 1000. There are 241 planets expected in the 2-minute cadence data, and lower limit of 977 planets in the FFIs.}
\end{figure*}

The product of these two terms can be seen in Figure~\ref{fig:yield}, for representative host stars and a observing baseline of $B=27.4$ days. To detect two transits, the probability is the geometric transit probability $R_{\star}/a$ until periods of $B/2$, while the probability of detecting a single transit peaks at $B$, with the transition for observing one versus two transits occurring between $\frac{B}{2}\le P\le B$. 

Each star observed by TESS will have an observing baseline of an integer multiple of $N\times27.4$ days where $1\le N\le13$. The amount of sky covered in each observing baseline is summarized in Table~\ref{tab:skycov}. The dominant baselines are 73.8\% of the mission covered for 27.4 days, 17.8\% for 54.8 days, and 3.5\% for 82.2 days.  There is an uptick at the ecliptic poles, which cover 2\% of the mission for 356 days. Each remaining observing baselines cover less than 1\% of the mission.

\begin{deluxetable}{cccc}[t]
    \startdata\\
        Days & Square Degrees & Sky Fraction & Mission Fraction\\
        \hline
        0 & 6023 & 14.6 & 0\\
        27.4 & 25989 & 63.0 & 73.8\\ 
        54.8 & 6270 & 15.2 & 17.8\\
        82.2 & 1238 & 3.0 & 3.5\\
        110 & 231 & 0.56 & 0.66\\
        137...301 & 578(total) & 1.4(total) & 1.6(total)\\
        329 & 215 & 0.52 & 0.6\\
        356 & 701 & 1.7 & 2.0\\
    \enddata\caption{\label{tab:skycov}Fraction of sky covered by various observing baselines.}
\end{deluxetable}

For the total number of stars observed by TESS, we use the TESS Candidate Target List-6 (CTL-6) provided online by \cite{stassun17}. The catalog has $\sim$4 million sources, with estimated host star masses, radii, and TESS magnitudes, the host star's coordinates, and and estimate of the blended flux from nearby stars that is expected to dilute the depth of transits. This method deviates from those used in simulations by \cite{sullivan15} and \cite{bouma17} as we do not use a Galactic model, but instead calculate our yields directly from the CTL-6. For each of the 4 million stars in the CTL-6, we use the estimated mass, radius, effective temperature, magnitude, and ecliptic latitude of the star.  As the the longitude of the first sector was not yet known, we use the ecliptic latitude to assign the number of sectors in which the star will be observed by TESS, and determines the total observing baseline. We do this by taking all stars above a given ecliptic latitude, such that the total area on sky is the same as described in Table~\ref{tab:skycov}. All observed stars are sorted by the CTL-6 priority, where the top 200,000 stars are classified as postage stamp (PS) stars, and the remaining stars are placed in the FFI sample. \cite{barclay18} have show that using only the CTL priority may result in an over-selection of target stars in the ecliptic poles relative to the true mission, however we remain agnostic as to the final selection strategy of the targeted stars and default to CTL-6 priority over speculation as to what the final selected stars will be. 

We use the planet occurrence rates of \cite{fressin13} for stars with $T_{\mathrm{eff}}\ge4000K$, and of \cite{dressing15} for stars with $T_{\mathrm{eff}}<4000K$. The planet occurrence rates are only complete to periods of $\sim100$ days, but we extrapolate these rates to periods of $>1000$ days to explore the probability of finding planets at longer periods. The assumed planet occurrence rates can be seen in Figure~\ref{fig:yield}. For each period and radius bin of the respective occurrence rates, we draw a radius and period from a random uniform logarithmic distribution in that bin. From the period of the planet, host star mass, and host star radius we calculate the geometric probability using Equation~\ref{eq:geo}. From the ecliptic latitude of the star and the period of the planet, we can calculate the probability of the object being observed for one, two, or more transits from Equation~\ref{eq:baseline}.  We note that, because the probability of detecting a single transit drops precipitously with period ($\propto P^{-5/3}$), changing our assumed form for the extrapolation  in period (within reason) is unlikely to change our results substantially.

We only calculate detections if the SNR is above 7.3. To calculate the SNR, we follow the formula used in \cite{bouma17}
\begin{eqnarray}
    SNR=\sqrt{N_{\mathrm{tr}}} \frac{\delta D} {\left(\frac{\sigma_{1\mathrm{hr}}^2}{T}+\sigma_{v}^{2}\right)^{1/2}},
\end{eqnarray}
using the number of transits $N_{\mathrm{tr}}=1$ for single-transits, the transit depth $\delta$, the dilution from background stars and contamination $D$ bounded from 0--1, the total noise per hour $\sigma_{1\mathrm{hr}}^2$ from CCD read noise, photon-counting noise, zodiacal noise, and a systematic 60 ppm hr$^{1/2}$ instrumental noise floor, the transit duration $T$, and an intrinsic variability term $\sigma_{v}$.

We use the host star's $T_{\mathrm{eff}}$ to assign an intrinsic stellar variability based on \cite{basri13} following the procedures of \cite{sullivan15} and \cite{bouma17}. Using the planet's period and radius, along with the host star's variability and magnitude, we calculate the SNR of each planet, in each period and radius bin, around every star in the sample. For those that have a $SNR>7.3$, we define the planet as being detected. We then sum the product of the geometric probability, probability of being observed, and planet occurrence rate, over all detected planets around all stars. The results are shown in Figure~\ref{fig:yield}. The total number of single-transit events is 1218, with 241 of the detections being found in the postage stamps. 201/241 have periods $>25$ days, and 19/241 have periods $>250$ days. Finally, we recover an estimate of the total integrated number of planets detected. As we integrate fractional probability over all stars, we only recover the total number of planets detected, and not the total number of host stars. As such we and cannot make any quantitative statements on the expected multiplicity of the systems, but one could assume that each star hosts only one planet.

\subsection{Demographics of Detected Single-Transits}

\begin{figure*}
    \centering
    \includegraphics[width=8cm]{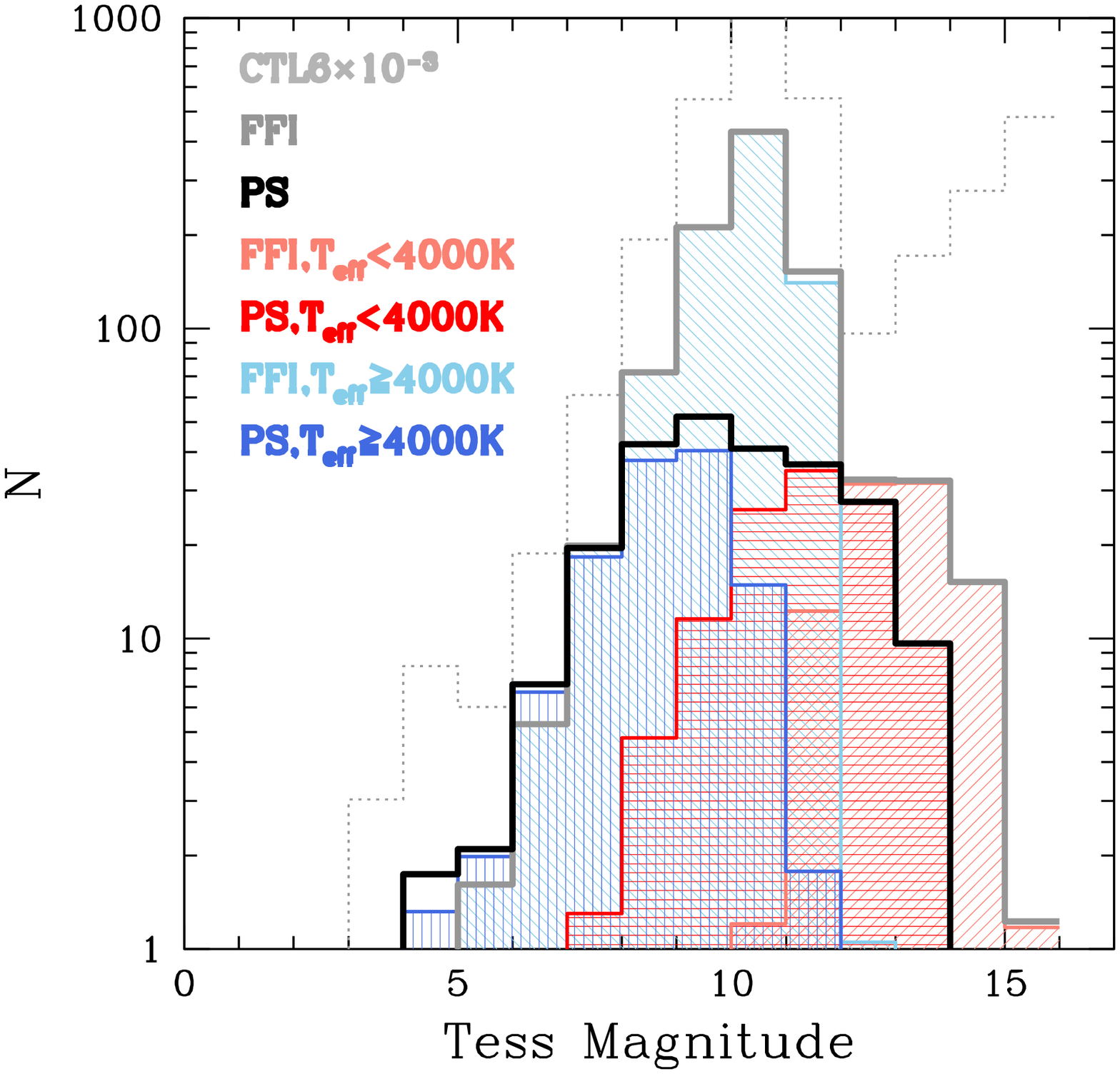}
    \includegraphics[width=8cm]{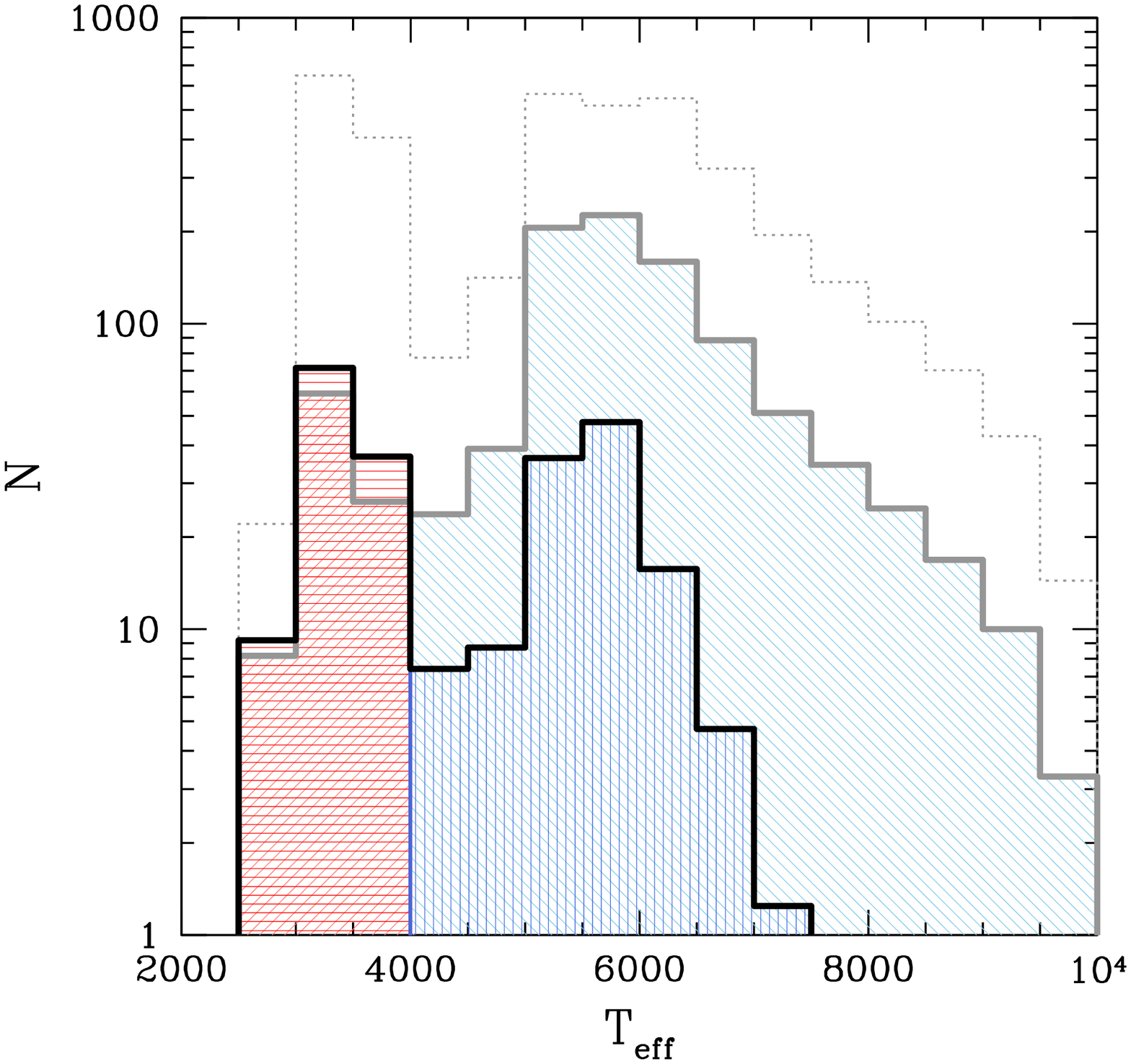}
    \includegraphics[width=8cm]{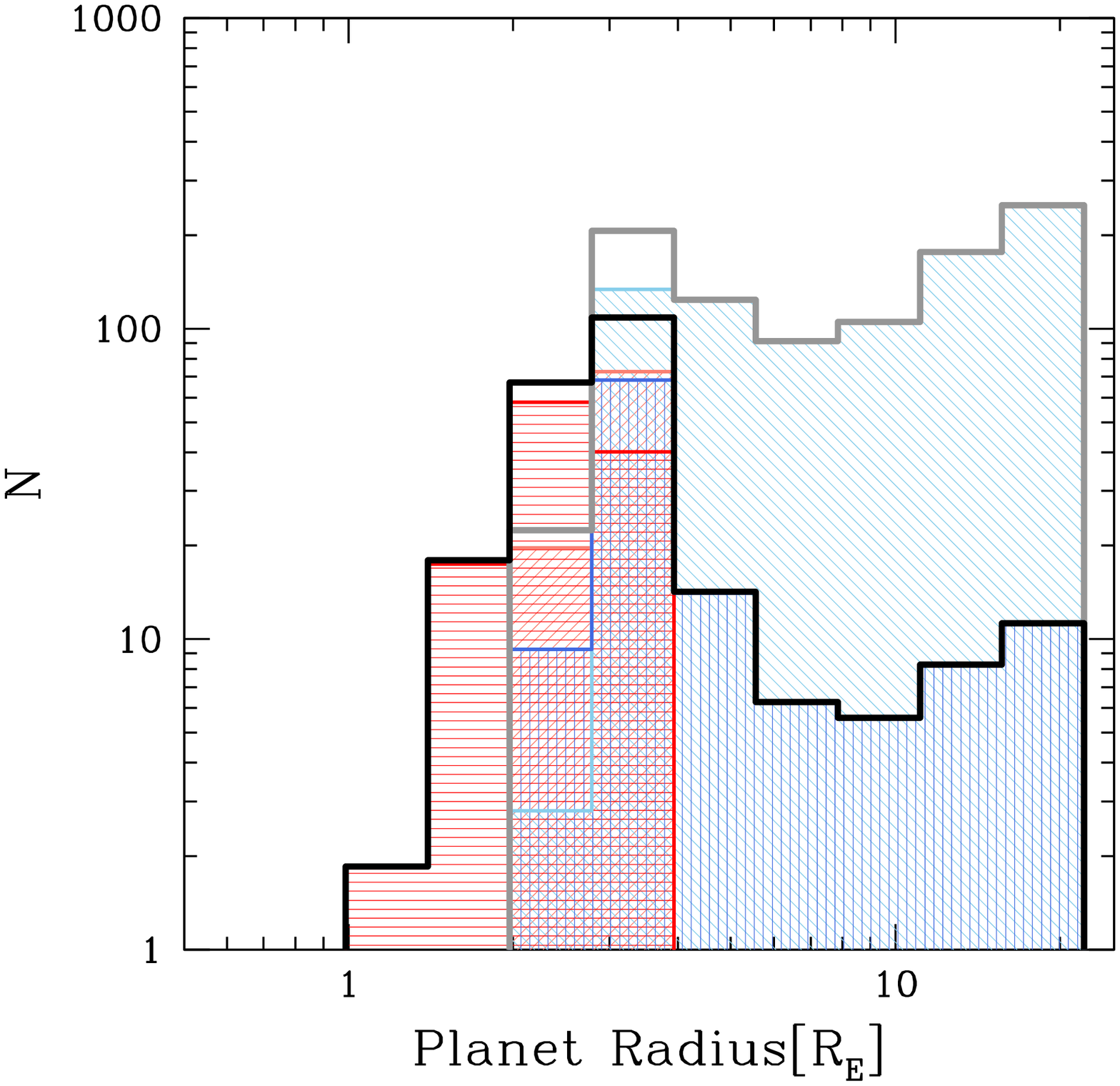}
    \includegraphics[width=8cm]{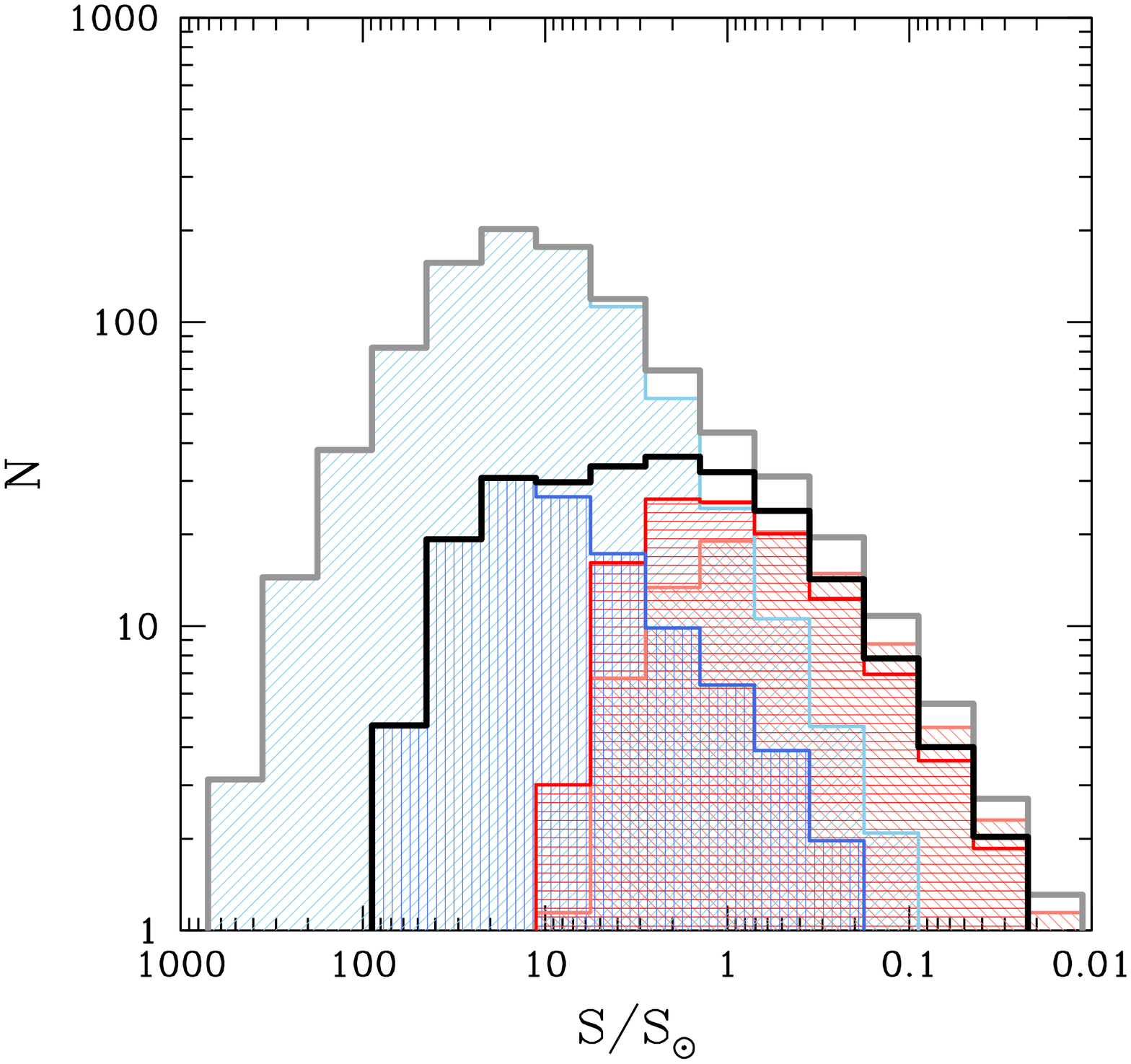}
    \caption{Number of expected single transit events by magnitude, $T_{\mathrm{eff}}$, planet radius, stellar insulation relative to the Sun. Dotted gray line is the demographics of the CTL-6 normalized to fit on this scale. The solid black and gray histograms are the total yield from the postage stamps and FFIs respectively. These are subdivided into $T_{\mathrm{eff}}\ge4000$~K (blue) and $T_{\mathrm{eff}}<4000$~K (red) star samples with the darker shades for postage stamps and the lighter shades for the FFIs.}
    \label{fig:demos}
\end{figure*}

We present the demographics of the detected planets. In Figure~\ref{fig:demos} we show the distribution of detections in host star magnitude, host star effective temperature, planet radius, and stellar insulation relative to the Earth. Of the 1218 expected single transits, 173 are around stars brighter than $T=10$ with 74 around postage stamp stars and 99 in the FFIs. In the postage stamps, the detected planets are split equally 118/124 among cool ($T_{\mathrm{eff}}<4000$~K) and warm ($T_{\mathrm{eff}}\ge4000$~K) stars, but the FFIs favor the warm stars with a 93/883 split between the cool/warm stars. We also find that 196 sub-Neptunes with $r_{p}<4R_{\oplus}$ will be detected as single-transits in the postage stamps, with an additional 230 detected in the FFIs of CTL-6 stars. All of the planets detected around cool stars have $r_{p}<4R_{\oplus}$ as there are no planets above $4R_{\oplus}$ in the \cite{dressing15} occurrence rates. \cite{clanton16} show that giant planets do indeed exist around cool stars at long periods, but are uncommon relative to small planets occurrence rates.

Fifty-six planets will be detected in the postage stamps with a stellar insulation within a factor of two of the Earth ($0.5\le S/S_{\odot}\le2$), and another 73 in the FFIs. Using the conservative habitable zone defined by \cite{kopparapu13}, we expect 34 habitable zone planets from the postage stamps, 29 are around cool stars, and another 45 habitable zone planets from the FFIs with 32 coming from cool stars. If we limit the habitable zone planets to terrestrial planets ($R\le1.5R_{\oplus}$) then we expect only 1 planet to be detected, which happens around a postage stamp star with $T_{\mathrm{eff}}\le4000$~K. In a recent simulation of the TESS yield from \cite{barclay18}, there were no planets found beyond $\approx85$ days which resulted in no planets being found in the habitable zone of FGK stars, where we find 5 in the postage stamps and 13 in the FFIs. It is worth noting that we differ from \cite{barclay18} not only in the number of habitable zone planets around stars earlier than M, but also in our target star selection criteria, extrapolation of planet occurrence rates, and in the definition of habitable zone.

\begin{figure}
    \centering
    \includegraphics[width=8cm]{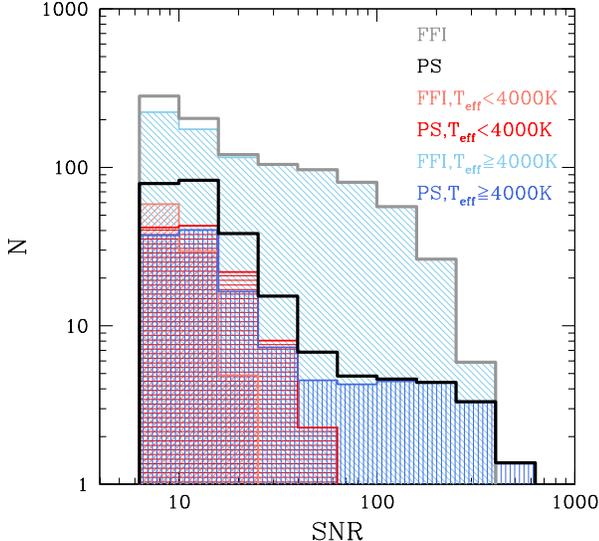}
    \caption{The number of single transit detections by SNR with a SNR threshold of $SNR=7.3$. Colors are the same as in Figure~\ref{fig:demos}. 162/241 postage stamp star detections have $SNR\ge10$, and 14/241 have $SNR\ge100$. 695/977 detections around FFI stars have $SNR\ge10$, and 90/977 have $SNR\ge100$.}
    \label{fig:snr}
\end{figure}

We adopt the same SNR threshold $SNR=7.3$ as used by both \cite{sullivan15} and \cite{bouma17} for multiple-transiting events. Given the added uncertainty of single-transit events (e.g. false positives) we follow \cite{barclay18} and also look at the distribution of SNR for all of the detections in Figure~\ref{fig:snr} for those wishing to adopt a more stringent SNR cut. We find that of the 241 postage stamp detections with $SNR\ge7.3$, 162 have $SNR\ge10$, and 14 have robust detections at $SNR\ge100$. Among the 977 detections around FFI stars with $SNR\ge7.3$, 695 have $SNR\ge10$, and 90 have $SNR\ge100$.

\section{Estimating the Period}
An important aspect of identifying the single-transit candidates is the ability to predict the time of future transits to confirm their ephemerides and schedule future observations. One can relate the observed properties of the light curve, the velocity of the planet assuming a circular orbit and zero impact parameter ($b=0$), and Kepler's third law (Eq.~\ref{eq:kepler}) to relate the period, stellar density, and the light curve observable quantities \citep{seager03,yee08,winn10}. Beginning with the velocity of the planet
\begin{eqnarray}
    v_{p}=\frac{2\pi a}{P}=\frac{2R_{\star}}{T_{\mathrm{dur},0}},
\end{eqnarray}
where $T_{\mathrm{dur},0}$ is the duration of the transit at $b=0$. One can relate $T_{\mathrm{dur},0}$ to the observable quantities $T_{\mathrm{dur}}$ and $\tau$, the measured duration of the transit and ingress/egress time, with
\begin{eqnarray}
    T_{\mathrm{dur}}&=&T_{\mathrm{dur},0}\sqrt{1-b^2}\\
    \tau&=&\frac{\sqrt{\delta} T_{\mathrm{dur},0}}{\sqrt{1-b^2}},
\end{eqnarray}
where $\sqrt{\delta}$ is the transit depth to arrive at:
\begin{eqnarray}\label{eq:perioddensity}
    P&=&\frac{G\pi^2}{3}\rho_*\left(\frac{T_{\mathrm{dur}}\tau}{\sqrt{\delta}}\right)^{3/2}
\end{eqnarray}
This leads to a degeneracy between the period and the host star density. \cite{seager03} showed that it is possible to estimate the period when the mass and radius of the host star is known. \cite{yee08} show that the fractional uncertainty in the period $\left(\sigma_{P}/P\right)$ will come from the fractional uncertainty in the density $\left(\sigma_{\rho}/\rho\right)$ and the fractional uncertainty on the period due to the TESS photometry $(\sigma_{P}/P)_{\mathrm{TESS}}$ added in quadrature:
\begin{eqnarray}
    \left(\frac{\sigma_P}{P}\right)^2=\left(\frac{\sigma_\rho}{\rho}\right)^2+\left(\frac{\sigma_P}{P}\right)^2_{\mathrm{TESS}}
\end{eqnarray}

\subsection{Uncertainty on the Period Due to the Stellar Density}

In order to place a constraint on the period, one also needs a constraint on the density of the host star. There are a few avenues that allow for an independent constraint on the density to better than 10\%, such that the photometry is the limiting factor of estimating the period of single-transit planets in TESS.

One method to constrain the density will be to first estimate the stellar radius $R_{\star}$ of the host star by combining a spectral energy distribution (SED) compiled from broadband photometric measurements or spectrophotometry (when available) with an estimate of the effective temperature $T_{\mathrm{eff}}$ of the star.  The effective temperature can be obtained from the SED itself or from high-resolution spectra of the host star.  By fitting the SED to a model stellar atmosphere, one can estimate the dereddened bolometric flux $F_{\star}$ and the extinction (adopting an extinction law).  With an estimate of $F_{\star}$ and $T_{\mathrm{eff}}$, one can then estimate the stellar angular diameter and thus physical radius of the star using a parallax $\pi$ from Gaia. The exact precision on $R_{\star}$ will depend on the quality of the parallax, SED, and spectra.

With the radius and the surface gravity of the host star $\log g$, the mass and therefore the density of the host star can be estimated.  There are a variety of ways one can estimate $\log g$. One can measure this quantity using gravity-sensitive lines in high-resolution spectra, although such spectroscopic estimates of $\log g$ can be relatively imprecise, and more importantly, inaccurate, particularly in some regions of parameter space. In some cases, granulation-based 'flicker' measurements can be used to obtain a more precise estimate of $\log g$ \citep{Bastien2013}, however this requires both high quality and relatively long-baseline photometry. 

In some cases, it may also be possible to constrain the density from astroseismology. \cite{kjeldsen95} show that the density scales with the measurable average large-frequency spacing $\langle\Delta\nu\rangle$ as $\rho_{\star}\propto\langle\Delta\nu\rangle^{2}$. In this case, only a $5\%$ measurement of $\langle\Delta\nu\rangle$ is required to place a constraint on the density to 10\%. As with flicker measurements, this requires high quality and relatively long baseline photometry.

For most cases, however, we expect that one will fit the radius determined as above, along with metallicity [Fe/H], $\log g$ and effective temperature from high-resolution spectra, to stellar isochrones, to determine precise (albeit model-dependent) estimates of the age, mass, and density of the star.  

The precision with which the density can be estimated for the star will ultimately depend on which method is used, and the quality of the data being used.  We will simply adopt a fiducial value of a 10\% precision on $\rho_{\star}$, but note that this may be optimistic in some cases.  

In the event that the single-transit planet is part of a multi-planet system with the host star hosting additional interior planets, it will be possible to obtain the density of the host star from transits of the inner planets if any of the inner planets have multiple transits detected and if there is an estimate of their eccentricity. From the period of the inner planets the density can be taken directly from Equation~\ref{eq:perioddensity} and applied as a constraint on the single-transit planet. \cite{ballard18} has shown that it is likely that multiple planet systems will be common in TESS, although many will be detected as single-planet systems because the additional planets may go undetected due to lack of SNR or because they only exhibit a single transit.

In the end, the exact precision will be determined by which observables are available, and their relative precision. The density constraints of individual stars will likely vary by orders of magnitude, but a 10\% precision is expected for many of the brighter, well characterized systems.

\subsection{Uncertainty on the Period Due to the Photometry}

The fractional uncertainty in the period due to the photometry $(\sigma_{P}/P)_{\mathrm{TESS}}$ is dominated by the ability to measure the ingress/egress time $\tau$. From Equation~9 in \cite{yee08}, we get that the fractional uncertainty in the period due to the photometric precision is
\begin{eqnarray}\label{eq:prec}
    \left(\frac{\sigma_P}{P}\right)^2_{\mathrm{TESS}}\approx\frac{9}{4}\left(\frac{\sigma_\tau}{\tau}\right)^2\approx\frac{1}{Q^2}\left(\frac{27T_{\mathrm{dur}}}{2\tau}\right),
\end{eqnarray}
and can be related to $Q$, the approximate total SNR of the transit, and the ratio of the transit duration $T_{\mathrm{dur}}$ to the ingress/egress time $\tau$, assuming $\tau\ll T_{\mathrm{dur}}$. A detailed investigation in to the details of the uncertainties in the observables can be found in \cite{carter08} and the Appendix of \cite{yee08}.

\begin{figure}
    \centering
    \includegraphics[width=8cm]{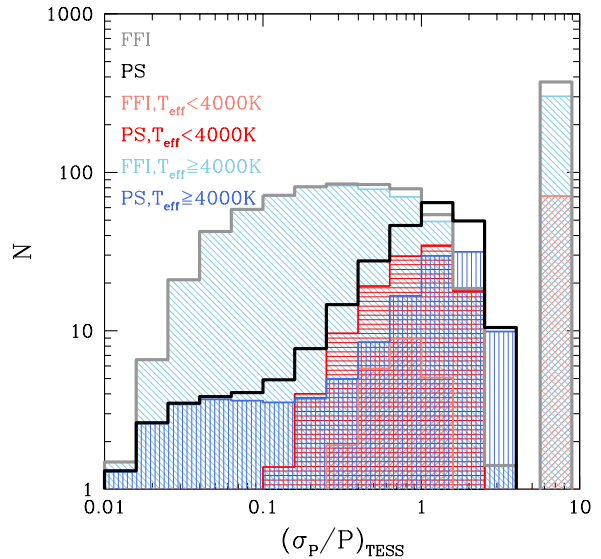}
    \caption{Number of expected single transit events by photometric uncertainty on the period. Colors are the same as in Figure~\ref{fig:demos}. The separate bin to the far right are the objects where the ingress/egress time was shorter than the exposure time and therefore only a upper limit can be place on the ingress/egress time.}
    \label{fig:precision}
\end{figure}

In Figure~\ref{fig:precision} we show the fractional uncertainty expected from single transits based on their photometry. 146/1218 planets will have a fractional uncertainty on the period to better than 10\%, with 16 coming from the postage stamps and 130 of those coming from the FFIs. It is worth noting that even in the event of a 1\% constraint on the period from photometry, the uncertainty on the density will likely dominate and limit the constraint on the inferred period. Another 72 planets will have a fractional uncertainty on the period of 10-15\%, with 5 around postage stamp stars and 72 coming from the FFIs. For cases where the ingress/egress time is shorter than the exposure time, we can only place an upper-limit on the ingress/egress time, and therefore lose the ability to constrain the period. This happens with 373 planets, all identified in the 30-minute cadence FFIs. The remaining 627 planets all have fractional uncertainties of greater than 15\%, where the approximation in Equation~\ref{eq:prec} breaks down and follow-up becomes difficult.

\subsection{Uncertainty on the Period Due to Eccentricity}

Up until now, we have assumed circular orbits for all of the estimates. In reality a number of these planets will likely have non-circular orbits. This will change the duration of the transit and will lead to an incorrect estimation of the period of the planet. \cite{yee08} show that the maximum and minimum deviation from the true period is given as:
\begin{equation}
    \left(\frac{\Delta P}{P}\right)_{\mathrm{min/max}}=\left(\frac{1+e}{1-e}\right)^{\pm3/2}
\end{equation}

\begin{figure}
    \centering
    \includegraphics[width=8cm]{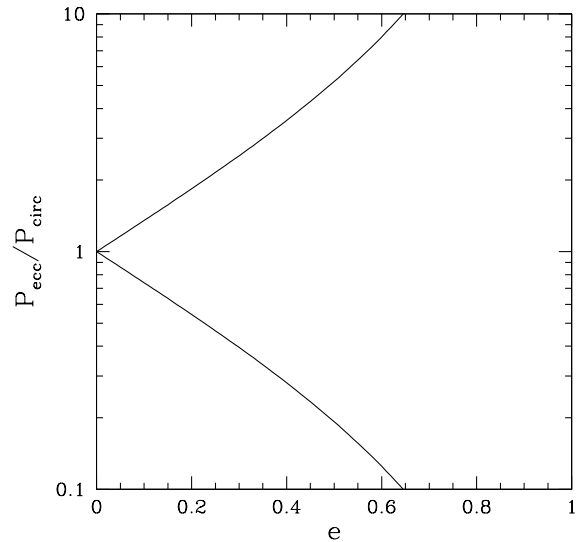}
    \caption{The maximum and minimum deviation from the true period under the assumption of a circular orbit. The spread is due to the change in planet velocity and transit duration when the planet transits at periastron versus apastron.}
    \label{fig:ecc}
\end{figure}

We show this in Figure~\ref{fig:ecc} where the minimum and maximum range of true periods of a planet with $e=0.1$ would be in the range of a factor of 0.74--1.35 of the assumed circular period. Assuming a median eccentricity of $e=0.17$ from the Beta distribution described in \cite{kipping13}, we get a typical range of periods from 0.59--1.69, or a $\approx50\%$ uncertainty in the period. This would imply that for non-circular systems, the $e\ne0$ uncertainties will limit over our ability to place a constraint using either density or the photometry. However, many of these systems will be in multiple planet systems \citep{ballard18}. \cite{zhu18} showed that Kepler planet systems become dynamically cooler as the number of planets in the system increases. This follows the results from \cite{xie16} that found the mean eccentricity of Kepler multi-planet systems of $e=0.04$ to be much lower than that of the single-planet systems $e=0.3$. For an eccentricity of $e=0.04$, the range of possible periods relative to circular is only 0.89--0.12, which is in line with the expected uncertainty from the density and photometry.

\section{Prospects for Follow-up}

\subsection{Recovery with Additional Photometry or Precovery in Archival Data}

\begin{figure}
    \centering
    \includegraphics[width=8cm]{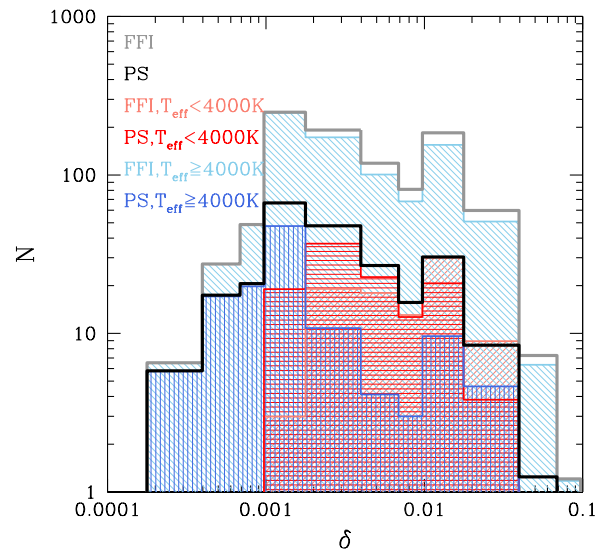}
    \caption{Expected single transit transit depths. Colors are the same as in Figure~\ref{fig:demos}. Of planets detected, 90\% and 24\% of the planets will have transit depths deeper than 0.1\% and 1\% respectively.}
    \label{fig:depths}
\end{figure}

After estimating the timing of a future transit, we also need to consider which of the single transit candidates will be observable from a typical ground based facility. An additional resource is to look for signals present in existing data sets given a known depth and approximate period. We present the distribution of the undiluted transit depths of single-transits in Figure~\ref{fig:depths}. 197/241 planets will be detectable at $\delta \ge 0.1\%$ in the postage stamps, with 894/976 of planets detectable around stars in the FFIs. Of these, 40 planets around postage stamp stars will have deep $\delta \ge 1\%$ transit depths, and 253 planets around stars in the FFIs.

\subsection{Expected Radial Velocity Signal}

\begin{figure}
    \centering
    \includegraphics[width=8cm]{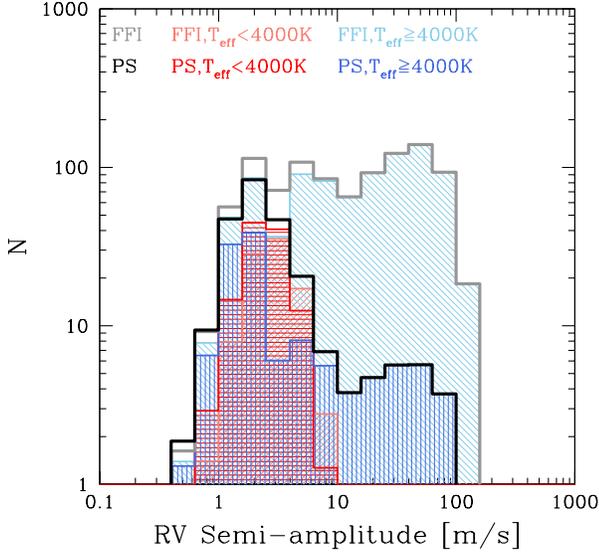}
    \caption{Expected single transit transit radial velocity signals. Colors are the same as in Figure~\ref{fig:demos}. Of planets detected, 98\% and 46\% of the planets will have RV signals greater than 1 and 10~m/s respectively.}
    \label{fig:rv}
\end{figure}

To estimate the expected radial velocity semi-amplitude $K$, we first assign a mass to each planet based on its radius. For planets with radii $<4.0~R_{\oplus}$ we use the planetary mass-radius relations from \cite{weiss14} and for planets with radii $\ge4.0~R_{\oplus}$ we use the planetary mass-radius relations from \cite{mordasini12}. After assigning a planetary mass $m_{p}$, we use the following equation to assign the radial velocity semi-amplitude $K$:
\begin{eqnarray}
    K=\frac{m_p}{(m_p+M_\star)^{2/3}}\left(\frac{P}{2\pi G}\right)^{-1/3}
\end{eqnarray}
The distribution of expected RV signals from the single transits is shown in Figure~\ref{fig:rv}. The majority of single-transit planets 1195/1218, will have RV semi-amplitudes detectable by modern RV instruments $K \ge 1$~m/s. 556/1218 will have $K \ge 10$~m/s and 20 will have $K \ge 100$~m/s.

\section{Comparison to Other Simulations}
\begin{figure}
    \centering
    \includegraphics[width=8cm]{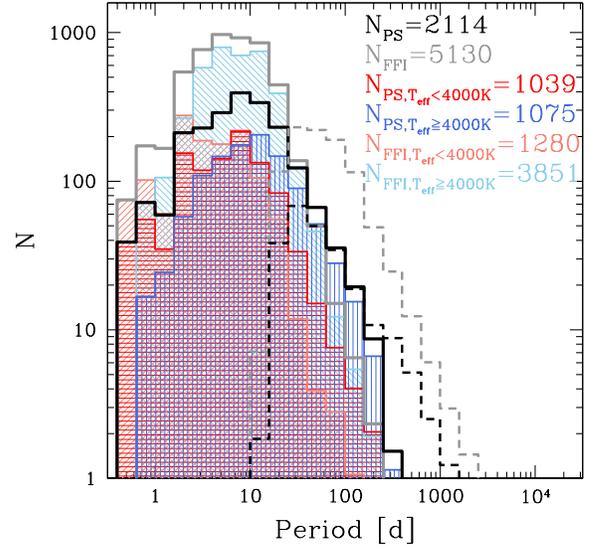}
    \caption{Expected yield from the TESS mission. Colors are the same as in Figure~\ref{fig:demos}. Single transit distribution from Figure~\ref{fig:yield} are shown in dashed lines. Single transits in the postage stamps match the TESS mission yield for planets with $25\le P \le250$~days, and dominate relative to the expected yield in the FFIs. Nearly all planets detected with $P\ge250$ come from the single-transits in both the postage stamps and FFIs.}
    \label{fig:mltyield}
\end{figure}

As no one has published yields from single transits from TESS simulations, we present the expected yield of the TESS mission proper (i.e. detected two or more transits) for our analysis as a way to compare and scale our results to previous studies. We perform the same analysis described in Section~\ref{sec:singles} to provide an updated estimate of the yield of the primary TESS mission while only considering the number of planets detected with two or more transits at a $SNR\ge7.3$. We find 2114 planets detected in the postage stamps and another 5130 in the FFIs around stars in the CTL-6. These can be seen in Figure~\ref{fig:mltyield}. Again, these numbers are incomplete fainter than $T>12$ and are lower limits for the FFIs. 255/2114 have periods $>25$ days, and only 2/2114 have periods $>250$ days in the postage stamps, while and 211/5130 have periods $>25$ days and $<1/5130$ have periods $>250$ days in the FFIs. We also show the yield from single transits as the dashed lines in Figure\ref{fig:mltyield} for the postage stamps (black) and FFIs (gray). We find that within postage stamps the single transits match the TESS mission yield beyond $\approx25$~days, and dominate relative to the expected yield in the FFIs. Beyond $\approx250$~days nearly all detections come from the single-transits in both the postage stamps and FFIs.

We find that our predicted yield from the postage stamps of 2114 represents a 20-25\% increase over the yield of both the \cite{sullivan15} (1734) and \cite{bouma17} (1670) estimates using a galactic model, and a 70\% increases over the more realistic \cite{barclay18} (1250) simulation. The \cite{sullivan15} yield predicts $>$20,000 planets detected in the FFIs, where we only detect 5130 planets in the FFIs, noting that we are incomplete and this is only a lower limit. \cite{bouma17} provides a lower limit of 3342 for the FFIs, which is consistent with our estimate. \cite{barclay18} found 1250 planets in the postage stamps, with another 3200 planets in the FFIs using the CTL-6 and TIC-6, with another 10,000 planets around stars faint stars not included in the CTL. We find that in general we over-estimate the number of planets in the postage stamps relative to other simulations, but are more consistent for the FFIs. It is worth noting that we have extrapolated our planet occurrence rates to much longer periods than all of the above simulations, and have a more simplistic target star selection criteria.

We were able to obtain rough numbers for the estimated number of single transit events from individual trials of other simulations via private communication. Again, we find 977 single-transits in the FFIs and 241 in the postage stamps in our work. From one trial from the \cite{sullivan15} simulation, we estimate $\approx1300$ single-transits in the FFIs and $\approx100$ single-transits in the postage stamps (\emph{P. Sullivan, private communication}). From one trial from the \cite{bouma17} simulation, we estimate  $\approx850$ single-transits in the FFIs and $\approx150$ single-transits in the postage stamps (\emph{L. Bouman, private communication}). Although unpublished, we find that the total number of single-transit events from each simulation, including our work, varies from 1000--1400 with each group disagreeing on the relative fraction found in the postage stamps versus the FFIs.

\section{Recommendations for Observations}

Given that nearly all of the 1218 single-transit events detectable in TESS will have either photometric (90\%) or radial velocity (98\%) signals measurable from current ground-based observatories, there will be more planets detected than could possibly be followed-up. Follow-up observers should coordinate to prioritize which planets will be targeted for follow-up observations, either by their ability to constrain the period, signal-to-noise ratio, or scientific merit. With 98\% of planets detectable in RV, that the RV measurements are required to determine the mass and planetary nature of the planets, and that RV measurements can help constrain the period, RV resources should be immediately allocated towards confirming single-transit events. RV will be crucial to the single transit detections in single-planet systems with poorly constrained eccentricities. Additionally, for those with photometric signals, searches in archival data and planned observations around the predicted next transit can be used to determine the period and constrain the timing of future transits.

\section{Conclusion}

The number of single-transit planets from TESS is expected to be an order of magnitude greater than those found in Kepler, with 241 single-transit planets detected in the postage stamps, and another 977 detected from the FFIs around stars brighter than $T=12$. Single transits require greater follow-up resources than the typical TESS planet, and there will be more single-transit planets signals than follow-up resources will be able to observe or confirm. This is despite the fact that 90\% and 98\% of all such planets detected will have photometric and RV signals respectively that will be observable from current ground-based observatories.

It is possible to predict future transits of single-transits by placing constraints on the light curve observables and on the density of the stellar host. The uncertainties from the density of the host star will be $\approx10\%$ in many cases, however only 10\% of the planets (146) will have photometry sufficient to provide constraints on the period to better than a 10\% due to uncertainties on the photometry assuming circular orbits. The uncertainty due to eccentric orbits will make constraining the true period difficult, but multi-planet systems represent the best systems to place constraints on both the stellar density and eccentricity.

Our single-transit yields predict a 80\% increase in the number of planets detected beyond 25 days compared to the TESS mission, and a factor of 12 increase in the yield for planets beyond 250 days. This includes 79 habitable zone planets and $\sim1$ terrestrial planet in the habitable zone. This opportunity to substantially augment the yield of the TESS mission should not be overlooked. However, given the``abundance of riches" represented by these single-transit events, we recommend community collaboration to make the most of these opportunities.  

\acknowledgements
\section{Acknowledgements}
We would like to thank Chelsea X. Huang for her feedback, helpful advice, and for sharing her yields with our group. We would like to thank Tom Barclay, Luke Bouma and Peter Sullivan for sharing their yields with our group. We would like to thank Daniel Stevens for his insight and feedback.

Work by S.V.Jr. is supported by the David G. Price Fellowship for Astronomical Instrumentation and by the National Science Foundation Graduate Research Fellowship under Grant No. DGE-1343012. D. Dragomir acknowledges support provided  by NASA through Hubble Fellowship grant HST-HF2-51372.001-A awarded by the Space Telescope Science Institute, which is operated by the Association of Universities for Research in Astronomy, Inc., for NASA, under contract NAS5-26555. Work by B.S.G. is supported by National Science Foundation CAREER Grant AST-1056524.

\end{document}